\documentclass[11pt,a4paper]{article}
\usepackage[a4paper,total={160mm,220mm}]{geometry}

\usepackage{amsmath,amssymb}
\usepackage{graphicx}
\usepackage{cite}
\usepackage{caption}
\usepackage{subcaption}
\graphicspath{{img/}}

\usepackage{braket}
\usepackage{hyperref}

\newcommand{\md}{\mathrm{d}}
\newcommand{\me}{\mathrm{e}}
\newcommand{\phiq}{\tilde{\varphi}}
\newcommand{\op}{\hat{\mathcal{O}}}

\renewcommand{\vec}[1]{\mathbf{#1}}

\begin{document}
\numberwithin{equation}{section}
\title{
\vspace*{-0.5cm}{\scriptsize \mbox{}\hfill IPPP/21/23}\\
\vspace{3.5cm}
\Large{\textbf{The Breakdown of Resummed Perturbation Theory\\at High Energies}
\vspace{0.5cm}}}

\author{Sebastian Schenk\\[2ex]
\small{\em Institute for Particle Physics Phenomenology, Durham University,} \\
\small{\em South Road, Durham DH1 3LE, United Kingdom}\\[0.8ex]}

\date{}
\maketitle

\begin{abstract}
\noindent
Calculations of high-energy processes involving the production of a large number of particles in weakly-coupled quantum field theories have previously signaled the need for novel non-perturbative behavior or even new physical phenomena.
In some scenarios, already tree-level computations may enter the regime of large-order perturbation theory and therefore require a careful investigation.
We demonstrate that in scalar quantum field theories with a unique global minimum, where suitably resummed perturbative expansions are expected to capture all relevant physical effects, perturbation theory may still suffer from severe shortcomings in the high-energy regime.
As an example, we consider the computation of multiparticle threshold amplitudes of the form $1 \to n$ in $\varphi^6$ theory with a positive mass term, and show that they may violate unitarity of the quantum theory for large $n$, even after the resummation of all leading-$n$ quantum corrections.
We further argue that this is a generic feature of scalar field theories with higher-order self-interactions beyond $\varphi^4$, thereby rendering the latter unique with respect to its high-energy behavior.
\end{abstract}

\newpage

\section{Introduction}
\label{sec:introduction}

The role of perturbation theory in weakly-coupled scalar quantum field theories has previously been challenged by computations of high-energy processes involving the production of a large number of particles.
High multiplicity threshold amplitudes of the form $1 \to n$ exhibit a factorial growth for large $n$ at tree level, as suggested by perturbative~\cite{Cornwall:1990hh,Goldberg:1990qk,Brown:1992ay,Voloshin:1992mz,Argyres:1992np,Smith:1992kz,Smith:1992rq} as well as semiclassical~\cite{Son:1995wz,Khoze:2017ifq,Khoze:2018kkz} calculations (see also~\cite{Khoze:2018mey} for a comprehensive review of the latter).
This is in strong contradiction to the requirement of unitarity for a consistent quantum theory, independent of its coupling.
Therefore, it may either signal an end of the perturbative regime, or even a more serious breakdown of the theory's description at high energies.

Phenomenologically, a rapid growth of multiparticle amplitudes is particularly crucial in scenarios involving a Standard Model Higgs boson~\cite{ATLAS:2012yve,CMS:2012qbp}.
This has lead to interesting ideas where yet unknown physical phenomena may be hiding from us~\cite{Voloshin:1992rr,Jaeckel:2014lya}.
In particular, a novel ``Higgsplosion" mechanism\footnote{The underlying nature of a quantum field theory featuring ``Higgsplosion" may be somewhat unconventional in terms of localizability and unitarity~\cite{Belyaev:2018mtd,Monin:2018cbi,Khoze:2018qhz}. For a recent assessment see also~\cite{Curko:2019dtu,Abu-Ajamieh:2020wqn}.} may provide for some interesting high-energy phenomenology~\cite{Khoze:2017tjt,Khoze:2017lft,Khoze:2017uga,Khoze:2018bwa}.
From a theoretical perspective, it has been argued that the problem of multiparticle production at high energies is, to some degree, intrinsically non-perturbative (see, e.g.,~\cite{Dine:2020ybn}).
That is, the tree-level computation of a high multiplicity amplitude may already be probing large orders of perturbation theory as $n \gtrsim 1 / \lambda$, therefore suffering from the well-known factorial growth of a divergent series.
The latter requires a careful resummation as compared to a \emph{naive} summation~\cite{Dyson:1952tj}.
For instance, for a scalar field theory with quartic coupling $\lambda \varphi^4$, it has been conjectured that the associated $1 \to n$ amplitude, in the double-scaling limit $\lambda \to 0$ and $n \to \infty$ with $\lambda n$ constant,\footnote{Recently, semiclassical expansions in parameters of the form $\lambda n$ have also proven useful in quantum theories for the investigation of physical quantities at large quantum numbers, see, e.g.,~\cite{Hellerman:2015nra,Monin:2016jmo,Arias-Tamargo:2019xld,Banerjee:2017fcx,Alvarez-Gaume:2016vff,Hellerman:2017efx,Hellerman:2018sjf,Badel:2019oxl,Badel:2019khk,Antipin:2020abu,Antipin:2020rdw,Cuomo:2020rgt,Antipin:2021akb,Dondi:2021buw,Cuomo:2021ygt,Hellerman:2021yqz,Hellerman:2021duh,Antipin:2021jiw,Antipin:2021rsh}.} is of exponential form~\cite{Voloshin:1992nu,Khlebnikov:1992af,Libanov:1994ug,Libanov:1995gh,Bezrukov:1995qh,Libanov:1996vq}, $\mathcal{A}(n) \sim \exp \left( F / \lambda \right)$.
An exact exponentiation of the multiparticle amplitude has later been established more rigorously in the quantum mechanical analogue of $\varphi^4$ theory, which corresponds to the quantum anharmonic oscillator~\cite{Jaeckel:2018ipq,Jaeckel:2018tdj}.
Crucially, in the double-scaling limit, a negative exponent, $F < 0$, renders the theory compatible with the unitarity requirement.
Indeed, there is strong evidence that this is the case for $\varphi^4$ theory with a positive mass term, not subject to a spontaneous symmetry breaking~\cite{Voloshin:1992nu,Smith:1992rq,Argyres:1993wz}.\footnote{For theories featuring a spontaneously broken symmetry the situation is more complicated, as the first quantum correction to the exponent is positive~\cite{Smith:1992rq,Khoze:2017ifq,Khoze:2018kkz}.}
Therefore, a suitable resummation of large-order perturbation theory may capture all relevant physical effects in order to restore unitarity of the quantum theory in the high-energy limit.
This is further supported by numerical studies~\cite{Demidov:2018czx}.
In fact, naively, we expect this to be generically true for any theory with a unique global minimum, where non-perturbative quantum effects, such as instantons, are absent.
In practice, the latter are known to spoil perturbative expansions by rendering them non (Borel) resummable (see, e.g.,~\cite{Brezin:1977gk,Bogomolny:1977ty,Bogomolny:1980ur,Stone:1977au,Achuthan:1988wh,Liang:1995zq}).

As more of a curse than a blessing, this work is supposed to add another peculiar observation to the problem of multiparticle production at high energies.
We aim to calculate $1 \to n$ amplitudes at the kinematic threshold in a general class of scalar quantum field theories featuring a unique global minimum.
As a first example, we explore a theory of a real scalar field with a sextic self-interaction $\varphi^6$.
In this setting, we calculate the $1 \to n$ amplitudes by means of the generating-field technique~\cite{Brown:1992ay}.
Here, all multiparticle amplitudes are generated by the matrix element $\bra{0} \varphi \ket{0}$, evaluated in the presence of a source.
We exploit that, beyond tree level, the leading-$n$ corrections to the amplitude correspond to quantum corrections to the matrix element~\cite{Voloshin:1992nu,Libanov:1994ug,Libanov:1995gh}.
By examining the singularity structure of the classical field solution, we derive these quantum corrections to large orders in perturbation theory.
This, in turn, enables us to resum all leading-$n$ corrections of the multiparticle threshold amplitude.
Surprisingly, we find that, in the large-$n$ regime, the $1 \to n$ amplitude is of the form
\begin{equation}
	\mathcal{A}(n) = \mathcal{A}_{\mathrm{tree}} (n) \cosh \left( \sqrt{\frac{F}{\lambda}} \right) \, ,
\end{equation}
where $F$ is a function of the combination $\lambda^2 n^4$ only.
This is, quite remarkably, a considerably more profound structure than the exponentiation of the identical object in $\varphi^4$ theory.
Crucially, in the double-scaling limit $n \to \infty$ and $\lambda \to 0$ with $\lambda n^2$ constant, the inherent structure of the amplitude appears to strongly contradict the unitarity requirement of the quantum theory.
In contrast to $\varphi^4$ theory, it is even independent of the precise form of the function $F$, as there will inevitably exist a contribution with an exponentially growing real part for large $n$ at weak coupling.
Therefore, even the resummation of all leading-$n$ quantum corrections to the multiparticle amplitude sustains the possibly dire consequences for the consistency of the quantum theory.

In addition, we also go beyond $\varphi^6$ theory and argue that for generic scalar field theories with higher-order self-interactions and a unique global minimum, the multiparticle amplitudes at the kinematic threshold are schematically determined by a sum over exponentials of complex roots of a yet unknown function.
This observation renders $\varphi^4$ theory phenomenologically and theoretically very special, as it is the only candidate allowing for a consistent implementation of unitarity.
It is remarkable and well unexpected that perturbation theory at large orders, even if suitably resummed, appears to suffer from severe shortcomings at high energies, as we will elucidate in detail in this article.

This work is structured as follows.
We start by reviewing the derivation of multiparticle amplitudes at the kinematic threshold from a matrix element of the field operator in Section~\ref{sec:MultiparticleAmplitudes}.
In Section~\ref{sec:QuantumCorrections} we demonstrate how quantum corrections to the matrix element can be systematically computed, order by order in a quantum-loop expansion.
We then argue in Section~\ref{sec:Resummation} that these quantum corrections coincide with the leading-$n$ corrections to the multiparticle threshold amplitudes.
Using this relation, we demonstrate how the large-order perturbative expansion can be resummed.
In Section~\ref{sec:ArbitraryPotentials}, we generalize our findings to scalar field theories with higher-order self-interactions.
Finally, in Section~\ref{sec:conclusions}, we give a brief summary and discussion of our results before we conclude.

\section{Multiparticle Amplitudes at the Kinematic Threshold}
\label{sec:MultiparticleAmplitudes}

In this work, we are interested in high-energy processes involving the production of a large number of quanta from only a few initial-state particles.
As a particular example, we consider a theory of a real scalar field with a sextic self-interaction in $2+1$ dimensions,
\begin{equation}
	S = \int \md^3 x \, \left( \frac{1}{2} \left( \partial \varphi \right)^2 - \frac{m^2}{2} \varphi^2 - \frac{\lambda}{6} \varphi^6 \right) \, ,
\label{eq:action}
\end{equation}
where the dimensionless coupling constant is positive, $\lambda > 0$, to render the theory stable.
We further choose $m^2 > 0$ in order to avoid additional complications arising from a spontaneous breaking of the internal $\mathbb{Z}_2$ symmetry of the field.
In this scenario, it is in principle possible that a highly-energetic (i.e.~off-shell) particle decays into many copies, $\varphi^{\ast} \to n \varphi$.
In particular, we are interested in the high-energy regime with a large number of final-state particles, $n \to \infty$.
At the kinematic threshold where all final-state particles are at rest, the scattering amplitude associated to this process can be elegantly derived from classical field equations~\cite{Brown:1992ay}.
As our work heavily relies on this method, let us briefly review the underlying formalism in the following.

In general, originating from the LSZ reduction formula~\cite{Lehmann:1954rq}, the scattering amplitude for a $1 \to n$ process in an arbitrary scalar quantum field theory can be written as~\cite{Brown:1992ay}
\begin{equation}
	\mathcal{A}(n) = \left[\prod_{a=1}^n \lim_{p_a^2 \to m^2} \int \md^{4} x_a \, \me^{i p_a x_a} \left(m^2 - p_a^2 \right) \frac{\delta}{\delta J \left(x_a\right)} \right] \bra{0} \varphi \ket{0} \bigg\rvert_{J=0} \, ,
\label{eq:AnLSZ}
\end{equation}
where the matrix element $\bra{0} \varphi \ket{0}$ is evaluated in presence of a source term, $J \varphi$.
In this sense, the vacuum expectation value of $\varphi$ generates all $1 \to n$ amplitudes through the response to an external source $J$.
At the kinematic threshold, where the momenta of all final-state particles vanish, we can assume the source to be spatially uniform and expand it into plane waves of frequency $\omega$,
\begin{equation}
	J(t) = J_0 \me^{i \omega t} \, .
\end{equation}
In this setting, the on-shell regime for the final-state particles can be implemented by sending $\omega \to m$.
Therefore, when going onto the mass shell the functional derivatives in~\eqref{eq:AnLSZ} can be replaced by (see also~\cite{Voloshin:1992nu})
\begin{equation}
	\lim_{p_a^2 \to m^2} \me^{i p_a x_a} \left(m^2 - p_a^2\right) \frac{\delta}{\delta J \left(x_a\right)} = \lim_{\omega^2 \to m^2} \me^{i \omega t} \left(m^2 - \omega^2 \right) \frac{\delta}{\delta J(t)} \equiv \lim_{\omega^2 \to m^2} \frac{\delta}{\delta z_\omega (t)} \, ,
\end{equation}
where we have defined\footnote{Note that, strictly speaking, we have made a simplification by omitting the standard analytic continuation to the complex momentum-plane, $m^2 \to m^2 - i \epsilon$. In principle, this has to be imposed in order to avoid the poles of the propagator occurring on the real axis. However, such simplification does not change our conclusions. For more details we refer the reader to~\cite{Brown:1992ay}.}
\begin{equation}
	z_\omega (t) = \frac{J_0}{m^2 - \omega^2} \me^{i \omega t} \, .
\end{equation}
Crucially, for the calculation of the multiparticle amplitude in the on-shell regime, $\omega \to m$, one can take the limit of vanishing source, $J_0 \to 0$, such that $z_\omega(t)$ remains finite at the same time, $z_\omega(t) \to z(t) = z_0 \exp\left(imt\right)$~\cite{Brown:1992ay}.
Therefore, within this formalism, all $1 \to n$ scalar field amplitudes at the kinematic threshold can be obtained from the matrix element $\bra{0} \varphi \ket{0}$ by~\cite{Brown:1992ay}
\begin{equation}
	\mathcal{A}(n) = \left. \frac{\partial^n}{\partial z^n} \bra{0} \varphi \ket{0} \right\rvert_{z=0} \, .
\label{eq:AnDerivative}
\end{equation}
In this sense, the vacuum expectation value of the field operator (evaluated in presence of the source) is the generating function of all multiparticle amplitudes at the kinematic threshold.
While this is an elegant way of writing the $1 \to n$ amplitudes, let us now demonstrate how they may be computed in practice.

\bigskip

In principle, the expression for the multiparticle amplitude in~\eqref{eq:AnDerivative} captures \emph{all} quantum effects of the theory through the vacuum expectation value of the field $\varphi$, in the presence of the source $J$.
While an exact expression for the latter is beyond our computational reach, we can instead try and reconstruct the response of the field to the source from a perturbative expansion in terms of Feynman diagrams.
At tree level, the matrix element $\bra{0} \varphi \ket{0}$ can then be replaced by the classical field $\varphi_0$, i.e.~by the solution to the classical equations of motion driven by the source,
\begin{equation}
	\left( \partial^2 + m^2 \right) \varphi_0 + \lambda \varphi_0^5 = J \, .
\end{equation}
Clearly, the classical field solution will explicitly depend on $J$.
At the kinematic threshold, the latter is simply a spatially-homogeneous plane wave, thereby allowing for a different approach to obtain the solution for $\varphi_0$.
As we have pointed out earlier, in the on-shell limit, $\omega \to m$, one can take the vanishing source, $J_0 \to 0$, such that $z_\omega(t) \to z(t) = z_0 \exp\left(imt\right)$ remains finite.
Therefore, in this regime, the classical field satisfies the equations of motion without a source~\cite{Brown:1992ay}
\begin{equation}
	\left( \partial^2 + m^2 \right) \varphi_0 + \lambda \varphi_0^5 = 0 \, ,
\end{equation}
however subject to the boundary condition that $\varphi_0$ approaches $z(t)$ for vanishing coupling, $\lambda \to 0$.
That is, the source term in the equations of motion has been replaced by an alternative boundary condition.
One can then solve the classical field equation exactly by performing a perturbative expansion to all orders in $\lambda$.
After a resummation of the latter, we find that the classical field has the form
\begin{equation}
	\varphi_0 (t) = \frac{z(t)}{\sqrt{1 - \frac{\lambda}{12 m^2} z(t)^4}} \quad \text{with} \quad z(t) = z_0 \me^{imt} \, .
\label{eq:phi0}
\end{equation}
Clearly, this solution satisfies the boundary condition $\varphi_0 \to z(t)$ for $\lambda \to 0$, as expected.
Therefore, to leading order in the loop expansion, we have derived the generating function of the multiparticle amplitudes at the kinematic threshold in a $\varphi^6$ theory with unbroken symmetry.
Finally, by means of the relation~\eqref{eq:AnDerivative}, we can determine the tree-level amplitude for a $1 \to n$ process.
It is given by
\begin{equation}
	\mathcal{A}_{\mathrm{tree}} (n) = \left. \frac{\partial^n}{\partial z^n} \varphi_0 \right\rvert_{z=0} = n! \binom{\frac{n-3}{4}}{\frac{n-1}{4}} \left( \frac{\lambda}{12 m^2} \right)^{\frac{n-1}{4}} \, ,
\label{eq:AnTree}
\end{equation}
where the number of final-state particles has to satisfy $n = 4k + 1$ with integer $k$.
Intuitively, in the language of Feynman diagrams, this is an immediate consequence of the sextic coupling of the theory, where each self-interaction of the field produces precisely four additional quanta in the scattering process.
Similarly, the above result --- which we have derived entirely by solving classical field equations --- is in exact agreement with results obtained from a recursive computation of Feynman diagrams~\cite{Argyres:1992np}.
In addition, we remark that it matches the expectations from its quantum mechanical analogue, which corresponds to transition amplitudes in the sextic anharmonic oscillator~\cite{Schenk:2019kmx}.

Similar to earlier works on the high-energy behavior of $\varphi^4$ theory (see, e.g.,~\cite{Cornwall:1990hh,Goldberg:1990qk,Brown:1992ay,Voloshin:1992mz,Argyres:1992np,Smith:1992kz,Smith:1992rq,Son:1995wz,Khoze:2017ifq,Khoze:2018kkz}), we observe that the tree-level $1 \to n$ amplitude for $\varphi^6$ theory given in~\eqref{eq:AnTree} grows factorially with the number of final-state particles.\footnote{Naively, the factorial growth is due to the rapidly growing number of Feynman diagrams, while lacking a destructive interference between all partial amplitudes contributing to the process. Instead, all diagrams add up coherently.}
Therefore, at least naively, the amplitude clearly violates unitarity in the regime $n \to \infty$, independent of the coupling $\lambda$.
While this seems to strongly contradict one of the fundamental principles of quantum physics, one should note that high-energy processes of the form $1 \to n$ may intrinsically be non-perturbative for large $n$, even in a weakly-coupled theory.
That is, here, a naive perturbative approach at tree level may be problematic for $n \gtrsim 1 / \lambda$, as, for large-order perturbation theory, the effective expansion parameter $\lambda n$ is strongly coupled in this regime (see, e.g.,~\cite{Khoze:2017ifq,Khoze:2018kkz}).
Nevertheless, in principle, we still expect a suitable resummation of large-order perturbation theory to yield a physically meaningful answer.
One might therefore wonder if this rapid growth is merely an artifact of entering the realm of large-order perturbation theory already at tree level.
If so, it should eventually be remedied by suitably including quantum corrections.
The generating-field technique we have described in this section is well suited to account for these quantum corrections to the multiparticle amplitude, which will involve combinations of powers of $\lambda$ and $n$.
We will explore this prospect in the remaining part of this work.

\section{Adding Quantum Corrections to the Threshold Amplitude}
\label{sec:QuantumCorrections}

The generating-field technique for computing multiparticle amplitudes at the kinematic threshold intrinsically accounts for all quantum effects present in the theory, since the matrix element $\bra{0} \varphi \ket{0}$ is the generating function for all $1 \to n$ processes.
For instance, in the case of $\varphi^4$ theory this has been pioneered in~\cite{Voloshin:1992nu,Smith:1992rq,Libanov:1994ug,Libanov:1995gh}.
Remarkably, in this scenario, it was shown that the leading quantum corrections to the amplitude are powers of the combination $\lambda n^2$ and indeed exponentiate, $\mathcal{A}(n) \sim \exp \left( \lambda n^2 \right)$.
Similarly, in this section, we aim to derive the leading quantum corrections to the multiparticle threshold amplitudes in $\varphi^6$ theory by closely following the methods developed in~\cite{Libanov:1994ug}.
We also note that a similar calculation has been briefly presented in~\cite{Libanov:1996vq}, albeit reaching a different conclusion.
In principle, this technique can also be used to include couplings to other (heavy) fields~\cite{Voloshin:2017flq}.

Generally, we start by considering the quantum fluctuations of the field around the classical background, i.e.~we decompose the scalar field into a classical and a quantum part,
\begin{equation}
	\varphi = \varphi_0 + \phiq \, .
\end{equation}
Using this decomposition, we then study the dynamics of the quantum field $\phiq$ in the classical background $\varphi_0$,
\begin{equation}
	S = \int \md^3 x \, \left( \frac{1}{2} \left( \partial \phiq \right)^2 - \frac{1}{2} \left(m^2 + 5 \lambda \varphi_0^4 \right) \phiq^2 - \frac{10}{3} \lambda \varphi_0^3 \phiq^3 - \frac{5}{2} \lambda \varphi_0^2 \phiq^4 - \lambda \varphi_0 \phiq^5 - \frac{\lambda}{6} \phiq^6 \right) \, .
\end{equation}
In order to simplify notation, we will set the bare mass of the field to unity for the remaining part of this work, $m^2 = 1$.
The dependence on the latter can later be reintroduced on dimensional grounds.
As the quantum fluctuations are now treated separately from the classical background field, the generating matrix element is similarly decomposed,
\begin{equation}
	\bra{0} \varphi \ket{0} = \varphi_0 + \bra{0} \phiq \ket{0} = \varphi_0 + \varphi_1 + \varphi_2 + \ldots \, .
\end{equation}
Here, $\varphi_k$ denotes the $k$-th quantum-loop contribution to the vacuum expectation value of $\varphi$, generated by the quantum fluctuations.
As these loop corrections enter the multiparticle amplitude linearly, we can systematically compute the latter order by order in a loop expansion,
\begin{equation}
	\mathcal{A}(n) = \left. \frac{\partial^n}{\partial z^n} \bra{0} \varphi \ket{0} \right\rvert_{z=0} = \mathcal{A}_0 + \mathcal{A}_1 + \mathcal{A}_2 + \ldots \, ,
\end{equation}
where the contribution from the $k$-th quantum loop is given by
\begin{equation}
	\mathcal{A}_k(n) = \left. \frac{\partial^n}{\partial z^n} \varphi_k \right\rvert_{z=0} \, .
\end{equation}
Our goal in the following paragraphs is to evaluate this expression explicitly.
That is, we aim to reconstruct the quantum contributions to the vacuum expectation value of the field operator in presence of the source, and then compute their derivatives with respect to the latter.

\subsection{One-loop contribution}

\begin{figure}[t]
	\centering
	\includegraphics[width=0.3\textwidth]{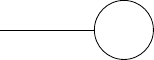}
	\caption{One-loop contribution to the vacuum expectation value of the quantum field $\phiq$.}
\label{fig:1-loop}
\end{figure}

By means of standard quantum field theory, the first quantum correction to the vacuum expectation value of $\varphi$ can be written in position space as
\begin{equation}
	\varphi_1 (x) = -10 \lambda \int \md^3 x^{\prime} \, D \left(x, x^{\prime}\right) \varphi_0^3 \left(x^{\prime}\right) D \left(x^{\prime}, x^{\prime}\right) \, ,
\label{eq:1-loop}
\end{equation}
where $D(x,y)$ denotes the Feynman propagator of the quantum field $\phiq$ and $\varphi_0$ is the classical background field given in~\eqref{eq:phi0}.
The corresponding Feynman diagram is illustrated in Fig.~\ref{fig:1-loop}.
Note that, here, we have implicitly considered the theory in a Euclidean spacetime for reasons that will become clear momentarily.
Clearly, in order to evaluate the expression for $\varphi_1 (x)$, we need to know the position-space representation of the propagator.
In general, the latter is the Green's function associated to the classical equations of motion of $\phiq$ in the non-interacting theory (i.e.~no self-interactions of $\phiq$),
\begin{equation}
	\left(\partial^2 + 1 + 5 \lambda \varphi_0^4 \right) D(x,y) = \delta^{(3)} (x-y) \, .
\label{eq:GreensFunction}
\end{equation}
Therefore, the overall strategy to obtain $D(x,y)$ is to invert the differential operator on the left-hand side of this equation.
Fortunately, the classical background field is spatially uniform and therefore only time-dependent, $\varphi_0 = \varphi_0(t)$, as given in~\eqref{eq:phi0}.
Hence, a particularly useful way to express the propagator is the mixed time-momentum representation~\cite{Voloshin:1992nu,Libanov:1994ug},
\begin{equation}
	D\left(x,x^{\prime}\right) = \int \frac{\md^2 \vec{p}}{\left(2\pi\right)^2} \, \me^{i \vec{p} \left( \vec{x} - \vec{x}^{\prime} \right)} D_{\vec{p}} \left(t, t^{\prime}\right) \, ,
\end{equation}
where we have split time and space coordinates, $x = \left(t, \vec{x}\right)$.
In fact, this representation of the propagator reduces~\eqref{eq:GreensFunction} to an ordinary differential equation of second order,
\begin{equation}
	\left( \partial_t^2 + \omega^2 + 5 \lambda \varphi_0^4 \right) D_{\vec{p}}\left( t,t^{\prime} \right) = \delta \left( t - t^{\prime} \right) \, .
\end{equation}
Here, we have defined $\omega$ to denote the energy of each momentum mode, $\omega^2 = \vec{p}^2 + 1$.
Again, in practice, in order to obtain $D_{\vec{p}}\left( t,t^{\prime} \right)$ we need to invert the differential operator on the left-hand side.
However, recalling that the classical background field $\varphi_0$ in~\eqref{eq:phi0} is essentially a complex number, we note that this operator is not Hermitian, thereby obstructing the inversion (see also~\cite{Voloshin:1992nu}).
To overcome this problem, we can do an analytic continuation in the time variable $t$.
That is, we perform a change of variable by writing
\begin{equation}
	\varphi_0^4 \left( \tau \right) = - \frac{12}{\lambda} \frac{u^4}{\left(1 + u^4\right)^2} = - \frac{3}{\lambda} \frac{1}{\cosh^2 (2\tau)} \, ,
\label{eq:phi04}
\end{equation}
where we have defined $u^4 \equiv \exp (4 \tau) = \left(- \lambda / 12 \right) z(t)^4$.
Naively, this can be understood as a Wick rotation from Minkowski time $t$ to Euclidean time $\tau$, followed by a constant shift proportional to the coupling of the theory.\footnote{The explicit relation between $\tau$ and $t$ is not important here, but it can be obtained by solving the algebraic equation $\exp \left( \tau \right) = \left(- \lambda / 12 \right)^{1/4} z(t)$ for $\tau$. This is also the reason why we have used the Feynman rules in Euclidean spacetime from the beginning.}
Importantly, with this choice of variables the operator $\varphi_0^4$ will be real for real time $\tau$ and we can write
\begin{equation}
	\left( -\partial_{\tau}^2 + \omega^2 - \frac{15}{\cosh^2 (2\tau)} \right) D_{\vec{p}}\left(\tau, \tau^{\prime}\right) = \delta \left(\tau - \tau^{\prime}\right) \, ,
\label{eq:EOMEuclidean}
\end{equation}
thereby enabling us to obtain a solution for the propagator in the following.

\bigskip

Before we continue, let us make a few important remarks about the structure of the classical background field $\varphi_0$ in terms of the Euclidean time variable $\tau$.
In general, the field is a multivalued function of $\tau$.
This is due to the fourth complex root introduced by the change of variable, implying that there are four equivalent solutions to~\eqref{eq:phi04} in total.
Similarly, there are four possible solutions of the defining relation between $\tau$ and $t$.
More importantly, however, we notice that the field is singular at $\tau_s = i \pi / 4$, subject to the expansion
\begin{equation}
	\varphi_0 (\tau) \simeq \left( - \frac{3}{\lambda} \right)^{\frac{1}{4}} \frac{1}{\sqrt{2}} \frac{1}{\sqrt{i\left(\tau - \frac{i\pi}{4}\right)}} + \ldots \, ,
\label{eq:singularity}
\end{equation}
where the dots represent terms regular at $\tau_s$ and we take the principal root of the $\tau$-dependent denominator.
Indeed, we also expect the quantum corrections to the classical field value to exhibit a singularity at $\tau_s = i \pi / 4$.
That is, schematically, in the vicinity of this singularity, the $k$-th quantum contribution is expected to be of the form~\cite{Libanov:1994ug}
\begin{equation}
	\varphi_k \simeq a_0 \varphi_0^{n_k} + a_1 \varphi_0^{n_k-1} + a_2 \varphi_0^{n_k-2} + \mathcal{O} \left(\varphi_0^{n_k-3}\right)  \, ,
\end{equation}
where $n_k$ is some positive number.
In fact, we will later show that $n_k = 4 k + 1$, thereby matching the number of scalar quanta that can be produced by an integer number of field self-interactions (cf.~Section~\ref{sec:MultiparticleAmplitudes}).
Importantly, we will also demonstrate that the leading-$n$ corrections to the $1 \to n$ multiparticle threshold amplitudes are in one-to-one correspondence with the leading singularities of the quantum corrections.
In other words, it is sufficient to determine the leading singularity structure of the quantum contributions to the matrix element in order to obtain the leading-$n$ contributions to the amplitudes.
We will make extensive use of this property in this work.

\bigskip

Let us now demonstrate how to obtain the propagator $D_{\vec{p}}\left(\tau, \tau^{\prime}\right)$ from the differential equation~\eqref{eq:EOMEuclidean}.
We first note that the formal solution is given by
\begin{equation}
	D_{\vec{p}} \left(\tau, \tau^{\prime}\right) = \frac{f_1(\tau)f_2(\tau^{\prime}) \theta \left(\tau - \tau^{\prime}\right) + f_2(\tau)f_1(\tau^{\prime}) \theta \left(\tau^{\prime} - \tau\right)}{W_{\vec{p}}} \, ,
\label{eq:propagator_formal}
\end{equation}
where the functions $f_1(\tau)$ and $f_2(\tau)$ are (linearly independent) solutions to the homogeneous equation and $W_{\vec{p}}$ is their Wronskian, $W_{\vec{p}} = f_1^{\prime} f_2 - f_1 f_2^{\prime}$.
In fact, the solutions to the homogeneous equation can be given exactly in terms of hypergeometric functions (see, e.g.,~\cite{Landau:1991wop}),
\begin{eqnarray}
	f_1 (\tau) &=& \frac{\me^{-\omega \tau}}{\Gamma \left( 1 + \frac{\omega}{2} \right)} \, {}_2F_1 \left( -\frac{3}{2} , \frac{5}{2} ; 1 + \frac{\omega}{2} ; \frac{1}{1 + \me^{4 \tau}} \right) \, ,\\
	f_2 (\tau) &=& \frac{\me^{\omega \tau}}{\Gamma \left( 1 + \frac{\omega}{2} \right)} \, {}_2F_1 \left( -\frac{3}{2} , \frac{5}{2} ; 1 + \frac{\omega}{2} ; \frac{1}{1 + \me^{-4 \tau}} \right) \, ,
\end{eqnarray}
where $f_1$ and $f_2$ are regular at $\tau \to -\infty$ and $\tau \to \infty$, respectively.
Using these solutions, we find that their Wronskian is also combination of hypergeometric functions,
\begin{align}
	W_{\vec{p}} =& - \frac{15}{8} \frac{1}{\Gamma\left(1+\frac{\omega}{2}\right)\Gamma\left(2+\frac{\omega}{2}\right)} \, {}_2F_1 \left(-\frac{3}{2}, -\frac{3}{2}+\frac{\omega}{2}; 1 + \frac{\omega}{2} ; -1\right) {}_2F_1 \left(-\frac{1}{2}, -\frac{3}{2}+\frac{\omega}{2}; 2 + \frac{\omega}{2} ; -1\right)  \nonumber \\
	&+ \frac{1}{4} \frac{\omega}{\Gamma\left(1+\frac{\omega}{2}\right)^2} \,  {}_2F_1 \left(-\frac{3}{2}, -\frac{3}{2}+\frac{\omega}{2}; 1 + \frac{\omega}{2} ; -1\right)^2
\end{align}
Intriguingly, one can check that the Green's function has poles (i.e.~the Wronskian vanishes) for two distinct energies, $\omega^2 = 1$ and $\omega^2 = 9$.
Naively, the LSZ reduction formula implies that on-shell tree amplitudes are non-vanishing only at these poles.
Therefore, scattering channels open up for only a certain class of on-shell diagrams, e.g.~in a $2 \to n$ process (for details, see~\cite{Voloshin:1992nu}).
In the rest frame of the final-state particles, the incoming particles have energy $\omega / 2$ and hence the poles only admit processes of the form $2 \to 2$ or $2 \to 6$, while higher values of $\omega$ lead to a vanishing tree-level threshold amplitude for $2 \to n$ processes with $n > 6$.
Indeed, this is in agreement with the analogous amplitudes in a $\varphi^4$ theory, where there also seems to exist a somewhat magical \emph{nullification} with all on-shell amplitudes, except for $2 \to 2$ and $2 \to 4$ scatterings, vanishing at the kinematic threshold~\cite{Voloshin:1992xb,Voloshin:1992nm,Argyres:1992un,Argyres:1993xa,Smith:1993hz}.
In this sense, the present example shares even more peculiarities with $\varphi^4$ theory.

\bigskip

Having obtained an exact expression for the Feynman propagator of the quantum fluctuations in the classical background, we can evaluate the one-loop contribution $\varphi_1(x)$ in~\eqref{eq:1-loop}, at least formally.
Nevertheless, in practice, the expressions are too involved in order to recover an exact result.
Instead, we make use of the singularity structure of the background field as illustrated in~\eqref{eq:singularity}.
We first note that the propagator is a Green's function of the operator $\op$ and therefore
\begin{equation}
	\op \varphi_1 (x) \equiv \left( \partial_x^2 + 1 + 5 \lambda \varphi_0^4(x) \right) \varphi_1(x) = -10 \lambda \varphi_0^3(x) D(x,x) \, .
\end{equation}
Here, the remaining propagator is evaluated at coinciding spacetime points.
As suggested in~\cite{Libanov:1994ug}, in order to evaluate this expression, we can focus on the leading singularities of the background field $\varphi_0$.
That is, if we are interested in the leading quantum contribution to the multiparticle amplitude, we only need to know the leading singularity of the loop-contribution $\varphi_1$ to the vacuum expectation value of $\varphi$.
In total, there are two contributions to the singularity structure of $\varphi_1$ in terms of the classical background field.
Clearly, the first is given by the vertex $\lambda \varphi_0^3$.
The second corresponds to the remaining propagator expression at coinciding points, $D(x,x)$.
That is, we can expand the latter around the singularity of the background, $\tau_s = i \pi / 4$, and obtain
\begin{equation}
	D (x, x) \simeq \lambda^{\frac{3}{2}} B \theta^2 \varphi_0^6 (\tau) \, ,
\end{equation}
where the equality is understood in terms of the leading singularity.
Here, $B$ is a numerical constant given by the momentum integration remnant of the propagator's Fourier transform,
\begin{equation}
	B = \frac{8\sqrt{3}}{9\pi} \int \frac{\md^2 \vec{p}}{\left(2\pi\right)^2} \, \frac{1}{\Gamma \left(\frac{5+\omega}{2} \right)^2} \frac{1}{W_{\vec{p}}} \, .
\end{equation}
Physically, $B$ inherits the divergences of loop-momentum integrals, which have to be carefully regularized.
Hence, it accounts for the renormalization properties of the theory.\footnote{In principle, our results are valid in an arbitrary number of dimensions, however, subject to the obvious complications arising from a renormalization procedure. For a detailed discussion, we refer the reader to~\cite{Voloshin:1992nu}.}
In addition, we have introduced the parameter $\theta$ to parametrize the complex phase of $\varphi_0$, corresponding to its four complex roots (see Eq.~\eqref{eq:singularity}).
In our example, this means that $\theta^4 = 1$, and hence it can take the values $\theta = \pm 1$ and $\theta = \pm i$.
Therefore, in the vicinity of the leading background singularity, the one-loop contribution to the matrix element satisfies
\begin{equation}
	\op \varphi_1 \simeq -10 \lambda^{\frac{5}{2}} B \theta^2 \varphi_0^9 \, .
\end{equation}
Similarly, it is straightforward to compute the inverse of the operator $\op$, by considering the leading singularity of the background field only.
That is, we further note that for some power $k$ of the field
\begin{equation}
	\op \varphi_0^k = -\frac{(k-3)(k+5)}{3} \lambda \varphi_0^{k+4} - \left(k^2 -1 \right) \varphi_0^k \, ,
\label{eq:recursive}
\end{equation}
and therefore, keeping only the leading singularity,
\begin{equation}
	\op^{-1} \varphi_0^k \simeq - \frac{3}{(k-7)(k+1)} \frac{1}{\lambda} \varphi_0^{k-4} \, .
\label{eq:recursive_leading}
\end{equation}
Using this property, we conclude that, in terms of the leading singularity, the one-loop quantum contribution to the vacuum expectation value of $\varphi$ reads
\begin{equation}
	\varphi_1 \simeq \frac{3}{2} \lambda^{\frac{3}{2}} B \theta^2 \varphi_0^5 \, .
\end{equation}
This is a crucial result, as it is in agreement with our earlier expectation that the quantum corrections share the singularity structure of the classical background field $\varphi_0$.
Moreover, we again note that this contribution depends on the complex phase of $\varphi_0$, which we have parametrized by $\theta$.
Indeed, as we will later see, this has important consequences for the resummation of the quantum contributions to the multiparticle threshold amplitudes.
Nevertheless, before discussing these in detail, let us first apply the same strategy to identify the two-loop quantum contribution to the matrix element.

\subsection{Two-loop contribution}

\begin{figure}[t]
	\centering
	\begin{subfigure}{0.24\textwidth}
		\centering
		\includegraphics[width=\columnwidth]{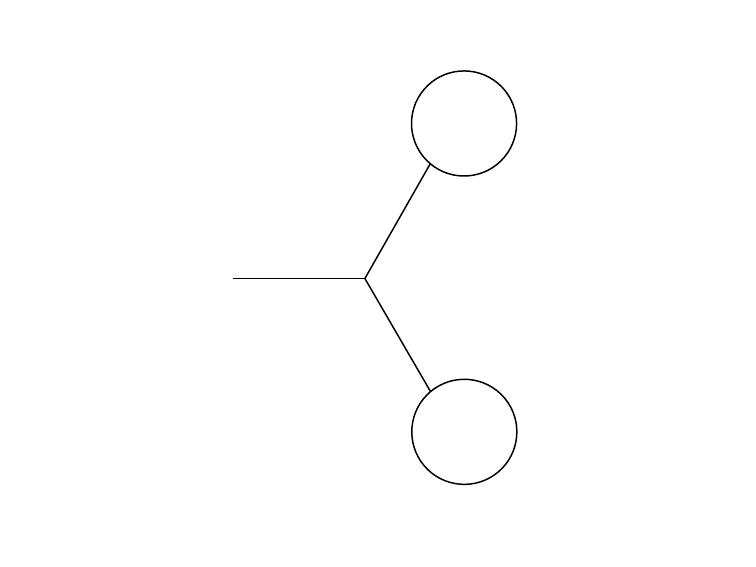}
		\caption*{$\varphi_2^a$}
	\end{subfigure}
	\begin{subfigure}{0.24\textwidth}
		\centering
		\includegraphics[width=\columnwidth]{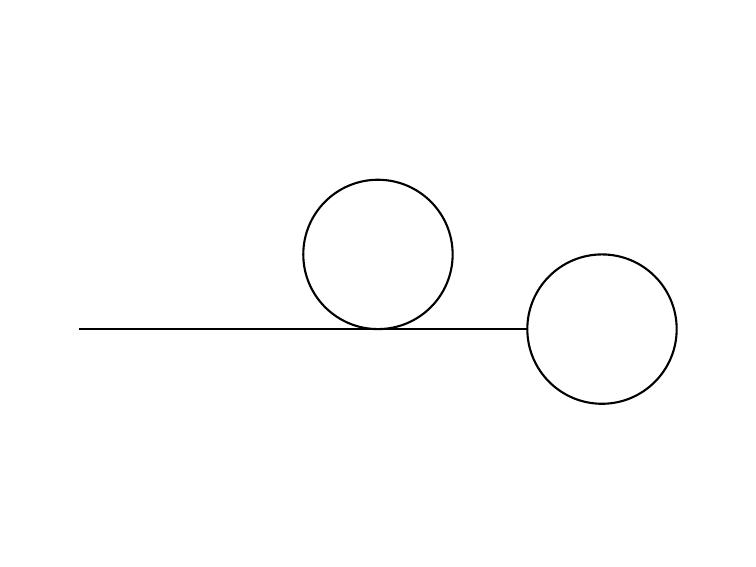}
		\caption*{$\varphi_2^b$}
	\end{subfigure}
	\begin{subfigure}{0.24\textwidth}
		\centering
		\includegraphics[width=\columnwidth]{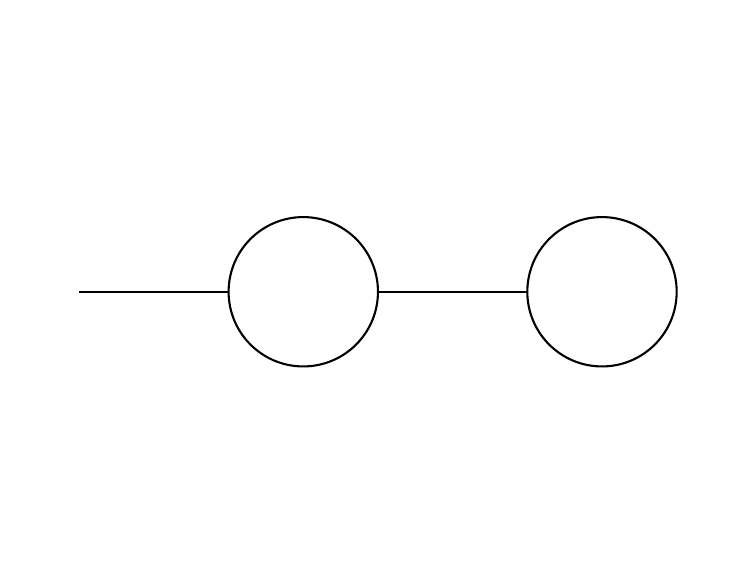}
		\caption*{$\varphi_2^c$}
	\end{subfigure}
	\begin{subfigure}{0.24\textwidth}
		\centering
		\includegraphics[width=\columnwidth]{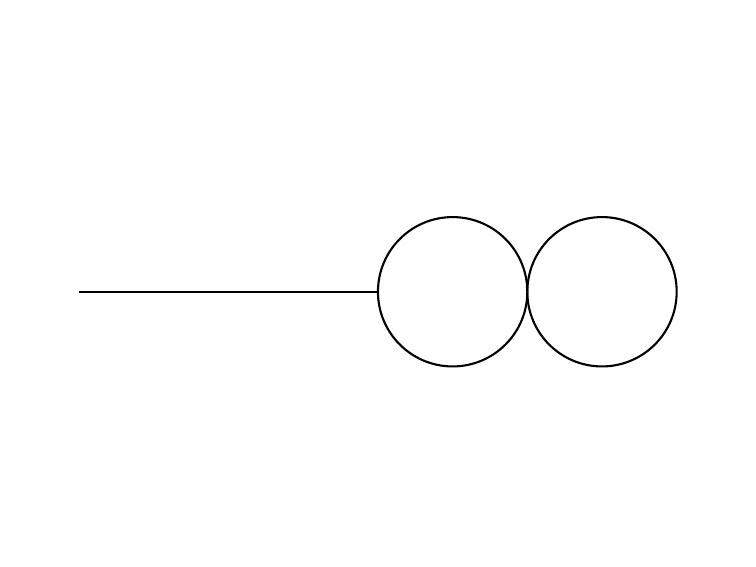}
		\caption*{$\varphi_2^d$}
	\end{subfigure}
	\begin{subfigure}{0.24\textwidth}
		\centering
		\includegraphics[width=\columnwidth]{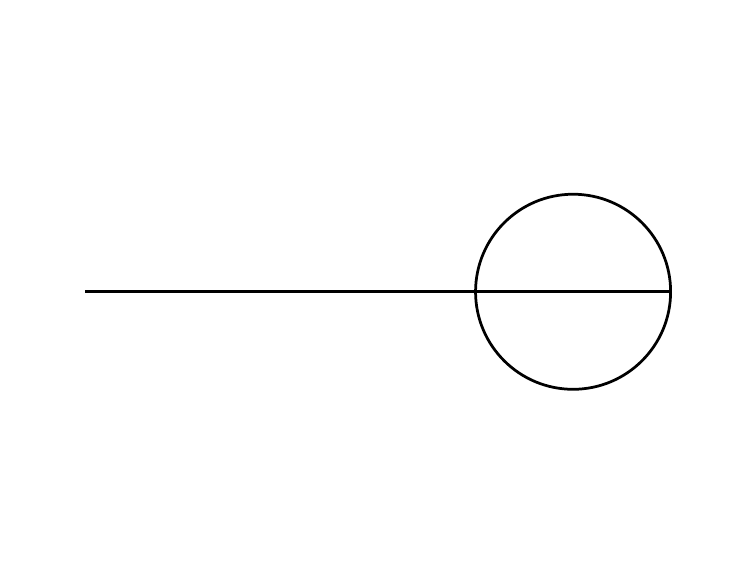}
		\caption*{$\varphi_2^e$}
	\end{subfigure}
	\begin{subfigure}{0.24\textwidth}
		\centering
		\includegraphics[width=\columnwidth]{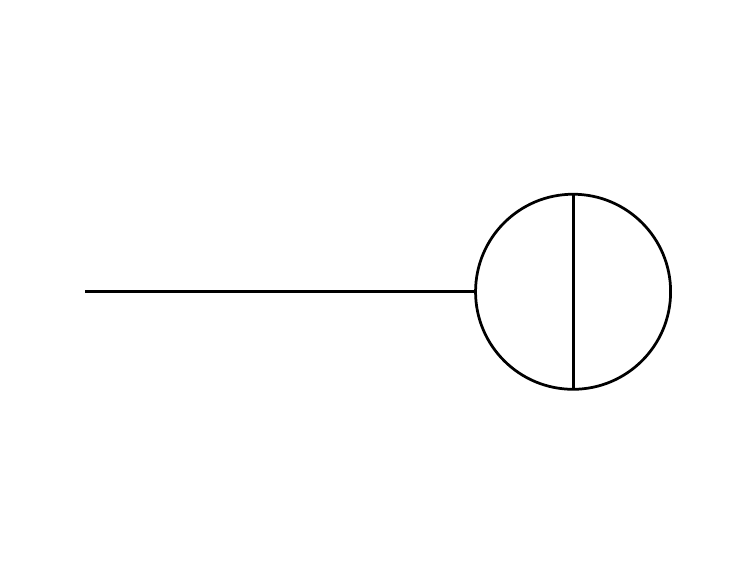}
		\caption*{$\varphi_2^f$}
	\end{subfigure}
	\begin{subfigure}{0.24\textwidth}
		\centering
		\includegraphics[width=\columnwidth]{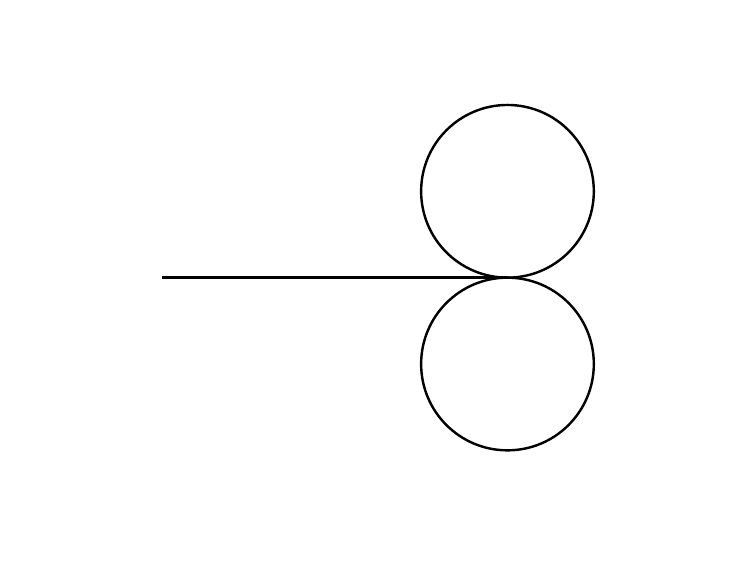}
		\caption*{$\varphi_2^g$}
	\end{subfigure}
	\caption{Two-loop contributions to the vacuum expectation value of the quantum field $\phiq$.}
\label{fig:2-loop}
\end{figure}

\begin{figure}[t]
	\centering
	\begin{subfigure}{0.24\textwidth}
		\centering
		\includegraphics[width=\columnwidth]{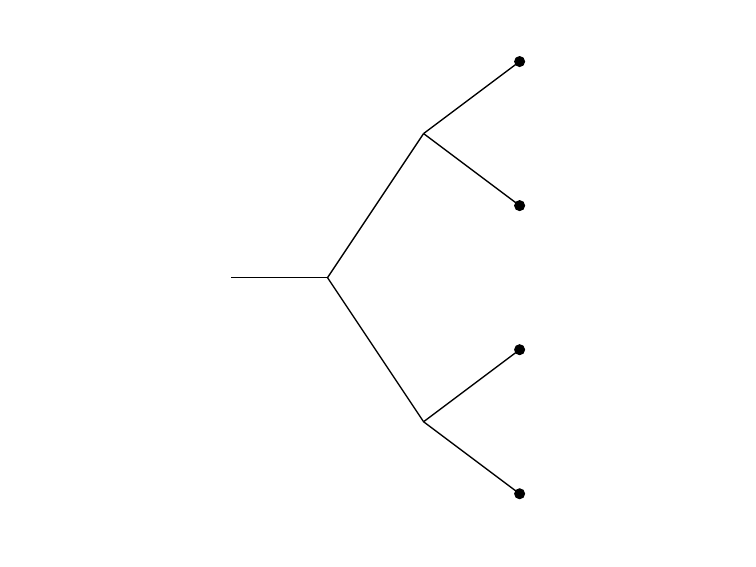}
		\caption*{$\varphi_2^a$}
	\end{subfigure}
	\begin{subfigure}{0.24\textwidth}
		\centering
		\includegraphics[width=\columnwidth]{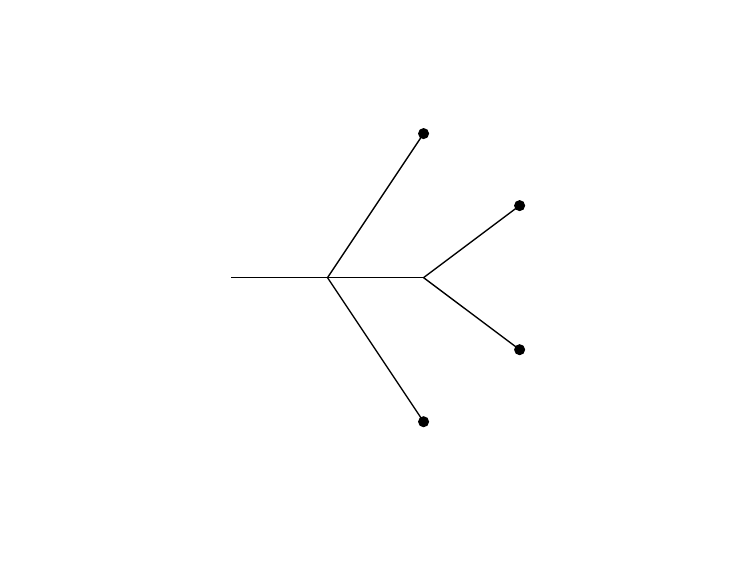}
		\caption*{$\varphi_2^b$}
	\end{subfigure}
	\begin{subfigure}{0.24\textwidth}
		\centering
		\includegraphics[width=\columnwidth]{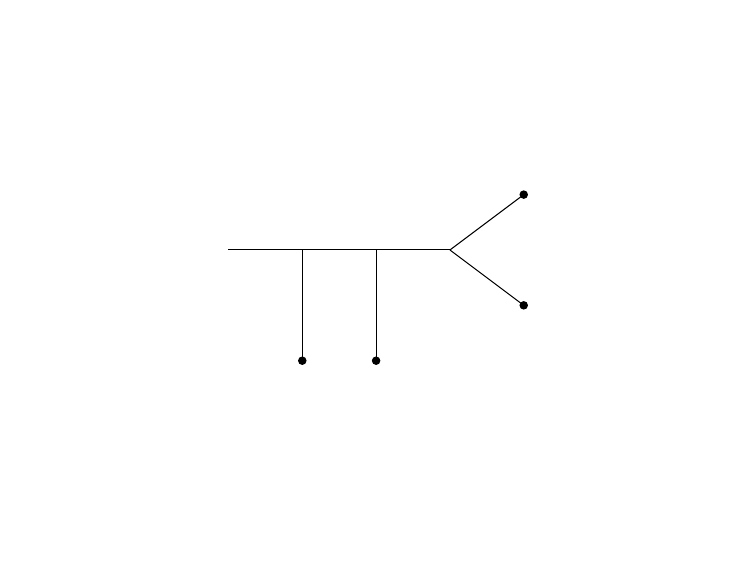}
		\caption*{$\varphi_2^c$}
	\end{subfigure}
	\begin{subfigure}{0.24\textwidth}
		\centering
		\includegraphics[width=\columnwidth]{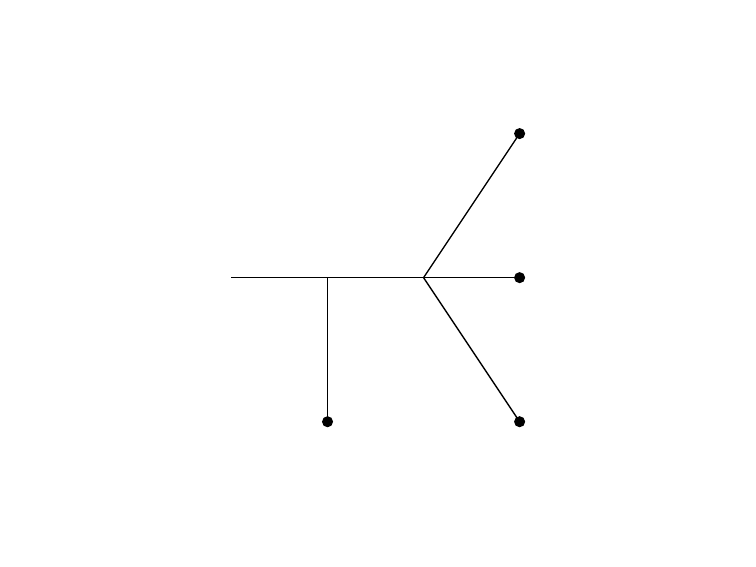}
		\caption*{$\varphi_2^d$}
	\end{subfigure}
	\begin{subfigure}{0.24\textwidth}
		\centering
		\includegraphics[width=\columnwidth]{2-loop_b_singular}
		\caption*{$\varphi_2^e$}
	\end{subfigure}
	\begin{subfigure}{0.24\textwidth}
		\centering
		\includegraphics[width=0.5\columnwidth]{2-loop_a_singular}
		\includegraphics[width=0.5\columnwidth]{2-loop_c_singular}
		\caption*{$\varphi_2^f$}
	\end{subfigure}
	\begin{subfigure}{0.24\textwidth}
		\centering
		\includegraphics[width=\columnwidth]{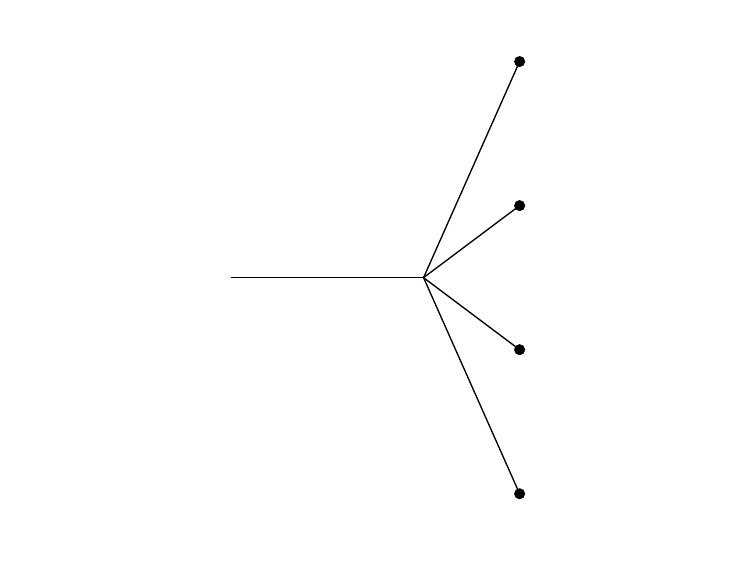}
		\caption*{$\varphi_2^g$}
	\end{subfigure}
	\caption{Two-loop contributions to the vacuum expectation value of the quantum field $\phiq$ represented by their leading singularities. The tree graphs emerge from cutting all internal propagators of the original Feynman diagrams, shown in Fig.~\ref{fig:2-loop}. Here, all external legs originating from a cut propagator are illustrated with a bullet point. Note that, for instance, the seemingly similar diagrams $\varphi_2^b$ and $\varphi_2^e$ come with a different symmetry factor.}
\label{fig:2-loop_singular}
\end{figure}

In general, we can repeat the analysis of the previous section in order to obtain the two-loop quantum contribution, $\varphi_2$, to the vacuum expectation value of $\varphi$.
In contrast to the one-loop case, there is an increased number of Feynman diagrams contributing to $\varphi_2$ which have to be taken into account.
These are illustrated in Fig.~\ref{fig:2-loop}.
In principle, we could go through the computation of every single Feynman diagram by performing all loop integrations explicitly.
However, remarkably, if we are only interested in the leading singularities of the quantum loops, the corresponding Feynman diagrams can be drastically simplified.
In fact, one can show that, in order to obtain the contributions from the leading singularities, one can evaluate tree graphs which descend from the loop diagrams~\cite{Libanov:1994ug}.
That is, for each loop diagram we have to cut all internal propagators, while accounting for all different possibilities and combinations of this procedure.
Effectively, this changes the symmetry factor of each diagram, as we will discuss in more detail momentarily.
In the emergent tree-level diagram, all internal lines then correspond to the inverse operator $\op^{-1}$ while external legs are associated with a factor $\sqrt{\lambda^{\frac{3}{2}} B \theta^2 \varphi_0^6}$, i.e.~a half-integer power of the Feynman propagator at coinciding spacetime points (see also~\cite{Libanov:1994ug}).
The descendent tree graphs are shown in Fig.~\ref{fig:2-loop_singular}, where for better visibility the external legs coming from a cut propagator are illustrated by a bullet point.
Indeed, this method is somewhat similar to well-known loop-tree dualities based on cutting propagators (see, e.g.,~\cite{Feynman:1963ax,Feynman:2000fh,Catani:2008xa}).

Let us demonstrate the drastic simplification through cutting internal propagators by an explicit calculation.
As an example, we focus on the two-loop contribution $\varphi_2^a$ shown in Fig.~\ref{fig:2-loop}.
In position space it is given by
\begin{equation}
	\varphi_2^a (x) = -10 \lambda \int \md^3 x^{\prime} \, D \left(x, x^{\prime}\right) \varphi_0^3\left(x^{\prime}\right) \varphi_1^2\left(x^{\prime}\right) \, .
\end{equation}
Applying the differential operator $\op$ and inserting the expression for $\varphi_1(x)$ we have derived previously, in the vicinity of the leading singularity of $\varphi_0$, we obtain
\begin{equation}
	\op \varphi_2^a = -10\lambda \varphi_0^3 \varphi_1^2 \simeq -\frac{45}{2} \lambda^{4} B^2 \theta^4 \varphi_0^{13} \, .
\end{equation}
Finally, using the recursive relation~\eqref{eq:recursive_leading}, we arrive at an expression for the two-loop contribution in terms of the leading singularity,
\begin{equation}
	\varphi_2^a \simeq \frac{45}{56} \lambda^3 B^2 \theta^4 \varphi_0^9 \, .
\end{equation}
Similarly, one can check that we get the same result by applying the prescription developed in~\cite{Libanov:1994ug} as briefly described above.
Let us point this out by referring to the corresponding tree-level Feynman diagram shown in Fig.~\ref{fig:2-loop_singular}.
Here, each internal line is associated with a factor $\op^{-1}$ and each bullet point comes with a factor $\sqrt{\lambda^{\frac{3}{2}} B \theta^2 \varphi_0^6}$.
Furthermore, there is only one possibility to cut all internal propagators in this two-loop diagram and hence, in this particular case, the symmetry factor is not changed with respect to the original diagram.
Therefore, within the novel set of emergent Feynman rules, the result may be written as
\begin{equation}
\begin{split}
	\varphi_2^a &\simeq \frac{(-20)^3}{8} \lambda^3 \op^{-1} \varphi_0^3 \left[ \op^{-1} \left( \varphi_0^3 \sqrt{\lambda^{\frac{3}{2}} B \theta^2 \varphi_0^6}^2 \right) \op^{-1} \left( \varphi_0^3 \sqrt{\lambda^{\frac{3}{2}} B \theta^2 \varphi_0^6}^2 \right) \right] \\
		&\simeq \frac{45}{56} \lambda^3 B^2 \theta^4 \varphi_0^9 \, ,
\end{split}
\end{equation}
where we have used~\eqref{eq:recursive_leading} in the second equality.
As expected, when only considering the leading singularity of the background field, we find an exact agreement between both approaches, i.e.~between the explicit loop computation and the method where only tree graphs have to be evaluated through cutting all internal propagators.

Before we continue, let us also comment on a relatively subtle simplification we have made above.
That is, when using the expression for $\varphi_1$ or, equivalently, the bullet point factor from a cut propagator, we have implicitly assumed that both elements enter the calculation with the same phase $\theta$.
However, in principle, both phases could be different, thereby leading to an additional interference term of the form $\varphi_2^a \propto \theta^2 \theta^{\prime 2} \varphi_0^9$.
Nevertheless, we have neglected this term for two reasons.
On the one hand, physically, it would clash with the continuity of the background field at coinciding points in spacetime.
On the other hand, as we will discuss in Section~\ref{sec:Resummation}, one has to include all possible phases in the coherent sum over Feynman diagrams, such that interference terms will give a vanishing contribution to the amplitude.
This is essentially because $\theta^4 = 1$.
Therefore, we only focus on terms without any interferences. 

In the following, we simply collect the results for the remaining Feynman diagrams shown in Fig.~\ref{fig:2-loop_singular}, which we have checked against an explicit loop computation.
Indeed, the methodology of~\cite{Libanov:1994ug} completely carries over to the present example, such that the quantum contributions read
\begin{equation}
\begin{split}
	\varphi_2^a \simeq \frac{45}{56} \lambda^3 B^2 \theta^4 \varphi_0^9 \, , \enspace  \varphi_2^b \simeq \frac{45}{28} \lambda^3 B^2 \theta^4 \varphi_0^9 \, , \enspace \varphi_2^c \simeq \frac{75}{56} \lambda^3 B^2 \theta^4 \varphi_0^9 \, , \enspace \varphi_2^d \simeq \frac{75}{56} \lambda^3 B^2 \theta^4 \varphi_0^9 \, , \\
	\varphi_2^e \simeq 2 \varphi_2^b \simeq \frac{45}{14} \lambda^3 B^2 \theta^4 \varphi_0^9 \, , \enspace \varphi_2^f \simeq 2 \varphi_2^a + 2 \varphi_2^c \simeq \frac{30}{7} \lambda^3 B^2 \theta^4 \varphi_0^9 \, , \enspace \varphi_2^g \simeq \frac{15}{28} \lambda^3 B^2 \theta^4 \varphi_0^9 \, .
\end{split}
\end{equation}
We observe that the symmetry factors for all diagrams obtained by cutting internal propagators at two-loop level come with an additional factor of $3$ as compared to a ``vanilla" tree-graph computation (see also~\cite{Libanov:1994ug}).
For instance, the symmetry factor of the contribution $\varphi_2^g$ is $1/8$, i.e.~precisely three times the naive tree-level factor we would apply to the computation of the corresponding tree graph in Fig.~\ref{fig:2-loop_singular}, $1/8 = 3/4!$.
This additional factor is also due to the different possibilities of cutting the internal lines, which have to be taken into account coherently.
For example, two-loop diagrams which are topologically different may lead to the same tree graph by the cutting procedure, as can for instance be seen from the contributions $\varphi_2^e$ and $\varphi_2^f$.
Finally, taking the leading singularity structure of all Feynman diagrams into account, the quantum correction to the generating matrix element at two-loop order is given by
\begin{equation}
	\varphi_2 \simeq \frac{105}{8} \lambda^3 B^2 \theta^4 \varphi_0^9 \, .
\end{equation}

Let us briefly summarize what we have found so far.
We have closely followed the methods developed in~\cite{Libanov:1994ug} in order to compute the quantum contributions to the matrix element $\bra{0} \varphi \ket{0}$, which generates all multiparticle threshold amplitudes according to~\eqref{eq:AnDerivative}.
By analyzing the leading singularities of the background field, we have been able to obtain the one-loop as well as the two-loop quantum correction to the matrix element,
\begin{equation}
	\bra{0} \varphi \ket{0} = \varphi_0 + \frac{3}{2} \lambda^{\frac{3}{2}} B \theta^{2} \varphi_0^5 + \frac{105}{8} \lambda^3 B^2 \theta^4 \varphi_0^9 + \ldots \, .
\label{eq:vev_2loops}
\end{equation}
Our result already indicates that, by evaluating tree graphs, this series can be continued to large orders in the quantum-loop expansion.
In turn, this allows us to access the large-order leading-$n$ contributions to the multiparticle threshold amplitudes.
In the following, we will argue that this can be done recursively and demonstrate that, in terms of the leading singularity, a complete resummation is within computational reach.

\section{Resummation of the Multiparticle Amplitude}
\label{sec:Resummation}

Quantum corrections to the vacuum expectation value of the field operator (in the presence of a source) can be computed order by order in a loop expansion.
In the previous section, we have explicitly performed the series expansion up to two-loop order, shown in~\eqref{eq:vev_2loops}.
In practice, the calculation is tremendously simplified by the fact that, in order to obtain the leading-$n$ corrections to the multiparticle threshold amplitudes, only the leading singularities of the background field have to be considered.
Using this observation, one can establish a one-to-one correspondence between intricate quantum-loop diagrams and much simpler tree graphs.
The generating matrix element $\bra{0} \varphi \ket{0}$ automatically encodes all quantum corrections to the multiparticle threshold amplitudes.
Consequently, we can gain valuable insights into the structure of $1 \to n$ amplitudes by determining the matrix element to large orders in perturbation theory.
In this section, we will demonstrate that this can be done explicitly and derive the multiparticle amplitude at threshold including all leading-$n$ corrections within the framework of the leading background singularity.

Let us first demonstrate how to extend the perturbative series for the vacuum expectation value of the field operator to large orders in the loop expansion.
Again, we will closely follow the arguments put forward in~\cite{Libanov:1994ug}.
In general, we can formally write down a series representation for the matrix element in the vicinity of the leading singularity of each quantum correction.
That is, with respect to the first three contributions shown in~\eqref{eq:vev_2loops}, we make the ansatz
\begin{equation}
	\bra{0} \varphi \ket{0} = \varphi_0 \sum_{k=0}^{\infty} d_k \left( \lambda^{\frac{3}{2}} B \theta^2 \varphi_0^4 \right)^k \, ,
\label{eq:vev_series}
\end{equation}
where the first coefficients of the series are $d_0 = 1$, $d_1 = 3/2$ and $d_2 = 105/8$.
In fact, there is an intuitive interpretation for these coefficients, $d_k$.
As discussed in Section~\ref{sec:QuantumCorrections}, at any loop order we can represent each Feynman diagram through a tree-level graph by cutting all internal propagators, while, at the same time, suitably changing the symmetry factor of the diagram.
This is illustrated in Fig.~\ref{fig:2-loop_singular}, where each bullet point corresponds to a factor of $\sqrt{\lambda^{\frac{3}{2}} B \theta^2 \varphi_0^6}$.
Therefore, the coefficient $d_k$ corresponds to the overall contribution from all tree-level graphs with $2k$ external legs, which emerge from cutting $k$ internal propagators.

The crucial observation made in~\cite{Libanov:1994ug} is that, equivalently, the emerging tree-level diagrams are generated by a classical field $\varphi_0$ that is shifted by the same half-integer power of the propagator,
\begin{equation}
	\varphi_0 \to \varphi_c = \varphi_0 + \sqrt{\lambda^{\frac{3}{2}} B \theta^2 \varphi_0^6} + \ldots \, .
\label{eq:condensate_shift}
\end{equation}
This allows us to make the following ansatz for the condensate
\begin{equation}
	\varphi_c = \varphi_0 \sum_{k=0}^{\infty} \alpha_k \left( \sqrt{\lambda^{\frac{3}{2}} B \theta^2 \varphi_0^4} \right)^k \, ,
\label{eq:condensate_ansatz}
\end{equation}
with $\alpha_0 = \alpha_1 = 1$, while \emph{a priori} the remaining coefficients $\alpha_k$ are unknown.
As the condensate generates all multiparticle threshold amplitudes via tree-level Feynman diagrams, it also satisfies the classical equations of motion,
\begin{equation}
	-\partial_{\tau}^2 \varphi_c + \lambda \varphi_c^5 = 0 \, ,
\end{equation}
where we have neglected the mass term for simplicity.\footnote{We note that the computation can also be performed with the mass term included. However, this does not lead to any different insights.}
This is the defining equation from which a general expression for the coefficients $\alpha_k$ can be derived.
For our purposes, it is sufficient to consider an expansion of the solution $\varphi_c$ around the leading singularity of the background field as shown in~\eqref{eq:singularity} (see also~\cite{Libanov:1994ug}).
One can then verify\footnote{Reversely, in order to arrive at this solution in practice, one would insert the original series ansatz~\eqref{eq:condensate_ansatz} into the equations of motion and solve it order by order in the coupling $\lambda$. The resulting perturbative expansion can then be resummed to obtain the solution $\varphi_c$.} that
\begin{equation}
	\varphi_c = \frac{\varphi_0}{\left(1 - \sqrt{4 \lambda^{\frac{3}{2}} B \theta^2 \varphi_0^4} \right)^{\frac{1}{2}}}
\end{equation}
is a solution to the classical equations of motion in the vicinity of the leading singularity of $\varphi_0$, that satisfies the condition~\eqref{eq:condensate_shift}.
The coefficients of the series expansion of the condensate are therefore given by
\begin{equation}
	\alpha_k = \frac{2^k}{\sqrt{\pi}} \frac{\Gamma\left(k+\frac{1}{2}\right)}{k!} \, .
\end{equation}
This is an important result, as we can now compare the Feynman diagrams generated by the field in both formulations at each order of the loop expansion.

In general, in a scenario where we shift the condensate, the symmetry factor for each tree-level diagram is different from the formulation where we cut all internal loops of the Feynman diagrams, thereby transforming it into a tree graph of the same order.
Let us briefly quantify this.
In the previous section, for instance, we have seen that the symmetry factor of each tree-graph is changed by a factor of $3$ at two-loop order and one can verify that it is multiplied by a factor of $15$ at three-loop order, respectively.
This precisely agrees with the results of~\cite{Libanov:1994ug} that entirely carry over to the present example.
That is, in general, the symmetry factor of each tree-graph originating from cutting $k$ propagators, i.e.~with $2k$ external legs or, equivalently, at $k$-th loop order, is multiplied by a factor $(2k)!/(2^k k!)$.
At the same time, the shifted condensate $\varphi_c$ generates tree-level Feynman diagrams with $k$ external legs at the same order in the series expansion.
Nevertheless, we expect both formulations to yield equivalent tree graphs.
Therefore, identifying the emergent diagrams by their number of external legs, the coefficients $d_k$ are given by
\begin{equation}
	d_k = \frac{\left(2k\right)!}{2^k k!} \alpha_{2k} = \frac{2^k}{\sqrt{\pi}} \frac{\Gamma \left(2k + \frac{1}{2}\right)}{k!} \, .
\label{eq:dk_coeff}
\end{equation}
Remarkably, we have just derived the large-order quantum corrections to the matrix element $\bra{0} \varphi \ket{0}$, which, in principle, enables us to resum the perturbative expansion.
Therefore, we can in turn access the large-order perturbation theory of the multiparticle amplitudes in question, as we will show next.

\bigskip

Earlier we have claimed that, in terms of the leading singularity, the quantum corrections to the matrix element correspond to the leading-$n$ contributions of the amplitude.
To see this, we first note that each quantum correction to the matrix element $\bra{0} \varphi \ket{0}$ scales with some power of the background field, $\varphi_0^{n_k}$.
Indeed, as we can see from~\eqref{eq:vev_series}, at the $k$-th loop order, the exponent is given by $n_k = 4k + 1$.
Consequently, using~\eqref{eq:AnDerivative}, the $k$-th quantum correction to the multiparticle amplitude in the large-$n$ regime then reads
\begin{equation}
	\mathcal{A}_k (n) \propto \left. \frac{\partial^n}{\partial z^n} \varphi_0^{4k+1} \right\rvert_{z=0} \sim \mathcal{A}_{\mathrm{tree}}(n) \left( \frac{\sqrt{\pi}}{\Gamma \left(2k + \frac{1}{2}\right)} \left( \frac{12}{\lambda}\right)^{k} n^{2k} + \ldots \right)  \quad \left( n \to \infty \right) \, .
\label{eq:Aloop}
\end{equation}
We observe that the leading-$n$ contribution at each loop order is dominated by the leading singularity of the classical background field $\varphi_0$, \emph{a posteriori} justifying our approach.
This is the reason why it is sufficient to only determine the leading singularity of each quantum correction $\varphi_k$ to the matrix element~\cite{Libanov:1994ug}.
Furthermore, using Eqs.~\eqref{eq:vev_series},~\eqref{eq:dk_coeff} and~\eqref{eq:Aloop}, we can finally resum all leading-$n$ contributions arising from quantum corrections to the multiparticle amplitudes in $\varphi^6$ theory to find
\begin{equation}
	\mathcal{A}(n) = \mathcal{A}_{\mathrm{tree}} (n) \sum_{k=0}^{\infty} \frac{1}{k!} \left(24 \sqrt{\lambda} B \theta^2 n^2\right)^k = \mathcal{A}_{\mathrm{tree}} (n) \exp \left( 24 \sqrt{\lambda} B \theta^2 n^2 \right) \, .
\end{equation}
Naively, similar to the case of $\varphi^4$ theory, we find an exact exponentiation of the leading quantum corrections to the multiparticle threshold amplitude at large $n$.
It is also in line with what has been claimed in~\cite{Libanov:1996vq}.

Although this seems to be a remarkable result in its own right, importantly, so far, we have still ignored the complex phase of the background field $\varphi_0$, which enters the expansion of the propagator around the leading singularity.
That is, in our parametrization, the above derivation has been performed for a fixed (but still arbitrary) phase $\theta$, everywhere.
As we have already pointed out in Section~\ref{sec:QuantumCorrections}, there may also be interference terms containing different phases from different propagators, e.g.~a two-loop term with $\theta^2 \theta^{\prime 2}$, contributing to the coherent sum of tree graphs.
However, while these naively clash with the continuity of the background field, in fact, they also vanish when all phases of the background field are taken into account.
This is essentially due to the fact that $\theta^4 = 1$ as discussed below.

As illustrated in~\eqref{eq:phi04}, the classical field has four equivalent branches in the complex plane due to the fourth complex root.
All complex branches of the field contribute equally to the leading singularity structure of the Feynman diagrams at each loop order.
Therefore, \emph{all} solutions have to be taken into account.
This means that we have to sum over all branches, i.e.~over all four complex roots of negative unity.
In our parametrization, $\theta^4 = 1$, the branches correspond to the possibilities $\theta = \pm 1, \pm i$.
Intriguingly, this in turn leads to an overall sum of two exponential functions, having similar exponents but of opposite sign.\footnote{Technically, because the series expansion of the threshold amplitude converges, we similarly could have first carried out the summation over all possible phases, observing that all mixed-phase terms drop out, before resumming the quantum corrections.}
Finally, the combination of both therefore yields a hyperbolic cosine,
\begin{equation}
	\mathcal{A}(n) = \mathcal{A}_{\mathrm{tree}} (n) \cosh \left( 24 \sqrt{\lambda} B n^2 \right) \, .
\label{eq:An_cosh_lambda}
\end{equation}
This is an extraordinary result, precisely matching the slightly more rigorous predictions from its quantum mechanical analogue~\cite{Schenk:2019kmx}.
Furthermore, it goes beyond the claims of~\cite{Libanov:1996vq} which did not include all complex branches of the classical background field.
The hyperbolic cosine also guarantees that a perturbative expansion of the multiparticle amplitude only contains integer powers of the coupling $\lambda$.
In order to align this expression with the exact exponentiation found in $\varphi^4$ theory, it is very suggestive to schematically write
\begin{equation}
	\mathcal{A}(n) = \mathcal{A}_{\mathrm{tree}} (n) \cosh \left( \sqrt{\frac{F}{\lambda}} \right) \, ,
\label{eq:An_cosh}
\end{equation}
where $F$ is a function of the combination $\lambda^2 n^4$.
This suggests that the $1 \to n$ threshold amplitudes in the large-$n$ regime are given by a sum over complex roots of some function $F$ divided by the coupling $\lambda$, indicating the inherent non-perturbative nature of the process.
To some degree, this can be understood as a rearrangement of the perturbative series, as the function $F$ now appears to be controlled by the effective parameter\footnote{Note that the effective expansion parameter looks slightly different here as compared to $\varphi^4$ theory, where the combination $\lambda n$ is taken. To some degree, this is an artifact of the definition of the scalar potential. For instance, to achieve a closer resemblance one could define the self-interaction term as $\lambda^2 \varphi^6$, which would lead to an effective expansion parameter $\lambda^2 n^2$. In this case, an expansion in $\lambda$ is also identical to an expansion in $\hbar$.} $\lambda n^2$.
It would be interesting to reconstruct this yet unknown function, for instance, in a $1/n$-expansion (see, e.g.,~\cite{Jaeckel:2018ipq,Jaeckel:2018tdj,Schenk:2019kmx}).
For a thorough investigation, more powerful resummation techniques may be needed (for some recent examples in $\varphi^4$ theory see, e.g.,~\cite{Serone:2016qog,Serone:2017nmd,Serone:2018gjo,Serone:2019szm,Sberveglieri:2019ccj,Romatschke:2019rjk,Heymans:2021rqo}).

\bigskip

In summary, we find that the leading-$n$ corrections to the multiparticle threshold amplitudes for a large number of final-state particles are determined by the leading singularities of the classical background field $\varphi_0$.
However, while in $\varphi^4$ theory the corrections resum into an exponential, here we find another exponential contribution with an exponent of opposite sign, combining into a hyperbolic cosine.
This remarkable structure is due to the four complex branches of the classical field which have to be taken into account equally.
In addition, as we will discuss in more detail in Section~\ref{sec:conclusions}, this may even have dire phenomenological consequences.
That is, in the double-scaling limit $n \to \infty$ and $\lambda \to 0$ with $\lambda n^2$ constant, the amplitude~\eqref{eq:An_cosh} grows without bounds even in a weakly-coupled theory.
Crucially, this is independent of the numerical constant $B$ or, in other words, even the precise form of the function $F$ --- one of the two exponential branches will always dominate the growth.
However, before exploring the consequences of this result, let us argue that this is by no means special to $\varphi^6$ theory but appears to be a generic feature of scalar quantum field theories with higher-order self-interactions.

\section{Higher-Order Self-Interactions}
\label{sec:ArbitraryPotentials}

The generating-field technique for the computation of multiparticle amplitudes at the kinematic threshold can straightforwardly be generalized to scalar quantum field theories with higher-order self-interactions,
\begin{equation}
	S = \int \md^d x \, \left( \frac{1}{2} \left( \partial \varphi \right)^2 - \frac{m^2}{2} \varphi^2 - \frac{\lambda}{k} \varphi^k \right) \, ,
\end{equation}
where, for simplicity, we choose all couplings to be positive, $m^2 > 0$ and $\lambda > 0$, and $k$ is an even integer with $k \geq 4$.
Let us again normalize the bare mass to $m^2 = 1$ in the following.
Note that, here, we are not concerned with the renormalization properties or physical content of this class of theories and therefore keep the number of spacetime dimensions arbitrary.
For theories with $k \geq 8$, with respect to a naive perturbative renormalization procedure we would need to consider a two-dimensional scenario, $d = 2$.
We also remark that the coupling constant $\lambda$ is not dimensionless in this setting, such that the weak-coupling limit $\lambda \to 0$ has to be taken with caution.
Nevertheless, in the present example, we are merely interested in the structural properties of the $1 \to n$ multiparticle amplitudes at the kinematic threshold.
In order to derive these, we can repeat the steps of Sections~\ref{sec:MultiparticleAmplitudes} and~\ref{sec:QuantumCorrections}.
That is, we first need to determine the classical background field in the presence of a source.
The latter satisfies the equations of motion
\begin{equation}
	\left( \partial^2 + 1 \right) \varphi_0 + \lambda \varphi_0^{k-1} = 0 \, ,
\end{equation}
with the boundary condition $\varphi_0 = z(t)$ for $\lambda \to 0$.
One can indeed verify that the solution to the classical field equation reads
\begin{equation}
	\varphi_0(t) = \frac{z(t)}{\left(1 - \frac{\lambda}{2k} z(t)^q\right)^{\frac{2}{q}}} \quad \text{with} \quad z(t) = z_0 \me^{it} \, ,
\end{equation}
where we have defined $q = k-2$.
Therefore, by means of the generating matrix element, the multiparticle threshold amplitude at tree level is given by
\begin{equation}
	\mathcal{A}_{\mathrm{tree}} (n) = n! \binom{\frac{n+1-q}{q}}{\frac{n-1}{q}} \left( \frac{\lambda}{2k} \right)^{\frac{n-1}{q}} \, ,
\end{equation}
agreeing with its recursively computed counterpart~\cite{Argyres:1992np}.
By computing the derivatives with respect to the source for different $n$, we also confirm that only final states with $n = qj+1$ for integer $j$ can be produced at tree level.
This in line with the expectations from a naive counting of vertices.

We can then proceed and consider the dynamics of the quantum fluctuations $\phiq$ of the field in the classical background $\varphi_0$, by decomposing $\varphi = \varphi_0 + \phiq$.
The action associated to the quantum part of the field can be written as
\begin{equation}
	S = \int \md^d x \, \left( \frac{1}{2} \left( \partial \phiq \right)^2 - \frac{1}{2} \phiq^2 - \frac{\lambda}{k} \sum_{i=2}^k \binom{k}{i}\varphi_0^{k-i} \phiq^i \right) \, .
\end{equation}
As discussed in detail in the previous sections, in order to obtain the contributions from the quantum fluctuations to the multiparticle amplitudes, we essentially need to determine the quantum corrections to the matrix element $\bra{0} \varphi \ket{0}$.
These can be identified order by order in a loop expansion.
A basic building block for this is the Feynman propagator $D(x,y)$, which satisfies
\begin{equation}
	\left(\partial^2 + 1 + \left(k-1\right) \lambda \varphi_0^{q} \right) D(x,y) = \delta^{(d)} (x-y) \, .
\label{eq:EOMGeneral}
\end{equation}
In principle, this Green's function can again be found by means of~\eqref{eq:propagator_formal}.
However, instead of repeating the lengthy computation explicitly, let us only sketch the main points.
In this sense, our derivation is not meant to be a fully rigorous calculation but rather a collection of consistency arguments.

In a first step, we determine the singularity structure of the classical background field.
A straightforward generalization of Section~\ref{sec:QuantumCorrections} reveals that, by a change of variable, defining $u^q \equiv \exp\left(q \tau \right) = -\left( \lambda / (2k)\right) z(t)^q$, we can write
\begin{equation}
	\varphi_0^q(\tau) = -\frac{2k}{\lambda} \frac{u^q}{\left(1+u^q\right)^2} = -\frac{k}{2} \frac{1}{\lambda} \frac{1}{\cosh^{2} \left( \frac{q\tau}{2}\right)} \, .
\end{equation}
Therefore, the positions of the singularities on the imaginary axis explicitly depend on the power of the self-interaction, $\tau_s = i \pi / q$.
In the vicinity of this singularity the classical background field is of the form
\begin{equation}
	\varphi_0 (\tau) \simeq \left(-\frac{k}{2} \frac{1}{\lambda}\right)^{\frac{1}{q}} \left(\frac{2}{q}\right)^{\frac{2}{q}}\frac{1}{\left[i \left(\tau - \frac{i\pi}{q} \right)\right]^{2/q}} + \ldots \, ,
\end{equation}
where the dots represent terms regular at $\tau_s = i \pi / q$ and we take the principal root of the $\tau$-dependent denominator.
Carefully note that we now have to take the $q$-th complex root of negative unity, i.e.~in total there will be $q$ equivalent solutions which have to be taken into account in the computation of the multiparticle amplitudes.
Again, this appears to be a straightforward generalization of the $\varphi^6$ example discussed before.

Similarly, the expansion around the leading singularity of the field can be used in order to express the propagator on the same footing.
Evaluated at coinciding spacetime points, in terms of the leading singularity, the propagator reads
\begin{equation}
	D(x,x) \simeq \lambda^{\frac{k}{q}} B \theta^2 \varphi_0^k (\tau) \, ,
\end{equation}
where $B$ is a numerical constant, involving the $(d-1)$-dimensional momentum integration of the Fourier transform.
Schematically, it can be written as
\begin{equation}
	B \propto \int \frac{\md^{d-1} \vec{p}}{\left(2\pi\right)^{d-1}} \, \frac{f(\omega)}{W_{\vec{p}}} \, ,
\end{equation}
for some function $f(\omega)$.
Here, $W_{\vec{p}}$ denotes the Wronskian associated to the solutions of the homogeneous equation of~\eqref{eq:EOMGeneral}.
Note that $B$ is not universal, i.e.~for each theory with self-interaction $\varphi^k$ we expect a different value.
Furthermore, the constant $\theta$ parametrizes each of the $q$ complex branches of the background field $\varphi_0$, such that~$\theta^q = 1$, yielding $q$ equivalent solutions in total.
Using the explicit form of the propagator evaluated at coinciding spacetime points, one can then argue that, by cutting all internal propagators at each loop order, the generating matrix element $\bra{0} \varphi \ket{0}$ will be of the form
\begin{equation}
	\bra{0} \varphi \ket{0} = \varphi_0 \sum_{i=0}^{\infty} d_i \left( \lambda^{\frac{k}{q}} B \theta^2 \varphi_0^q \right)^i \, .
\label{eq:vev_series_general}
\end{equation}
In a next step, applying the strategy of Section~\ref{sec:Resummation}, one can derive the exact form of the coefficients $d_i$, for each power of the theory $k$, which, however, we will not present here explicitly.
Instead, we immediately translate the generating matrix element into the leading-$n$ behavior for the multiparticle threshold amplitude at large $n$.
Remarkably, for an arbitrary but fixed complex branch of the background field, we again find an exact exponentiation of the leading-$n$ corrections,
\begin{equation}
	\mathcal{A}(n) = \mathcal{A}_{\mathrm{tree}} (n) \exp \left( \lambda^{\frac{2}{q}} B \theta^2 n^2 \right) \, ,
\end{equation}
where, for simplicity, we have omitted numerical constants in the exponent.
We again stress that this result is a generalization of the previous $\varphi^6$ example rather than a rigorous derivation.
Nevertheless, for a fixed phase, our result is consistent with earlier claims on multiparticle amplitudes in general scalar quantum field theories~\cite{Libanov:1996vq}.

We now observe an extraordinary structure emerge, once all $q$ complex branches of the classical field are taken into account.
By construction, these satisfy $\theta^q = 1$ and therefore, as $q$ is always even, will typically arrange in pairs of opposite sign in the complex plane.
This effectively reduces the number of branches by half to $q/2$ different contributions in total (due to the appearance of $\theta^2$ in the exponent of the amplitude).
Therefore, in summary, the leading-$n$ resummation of the multiparticle threshold amplitudes in $\varphi^k$ theory can be schematically written as
\begin{equation}
	\mathcal{A}(n) = \mathcal{A}_{\mathrm{tree}} (n) \sum_{\frac{q}{2} \, \mathrm{roots}} \exp \left( \sqrt[\frac{q}{2}]{\frac{F}{\lambda}} \right) \, .
\end{equation}
Here, the $F$ is a function of the combination $\lambda^2 n^q$, $F = F \left(\lambda^2 n^q\right)$.
That is, the $1 \to n$ multiparticle amplitudes in the large-$n$ regime involve a sum over $q/2$ complex roots of some function $F$, providing for a remarkable structure.
Ultimately, these complex roots descend from the multivalued classical background field $\varphi_0$.
Strikingly, an identical structure of the multiparticle amplitudes has been found earlier in the quantum mechanical analogue of $\varphi^k$ theory~\cite{Schenk:2019kmx}.
Note that, similar to the constant $B$, the function $F$ is not universal.
Instead, the precise form will depend on the power of the self-interaction.
We also remark that, similar to the previous section, our result goes far beyond the claims of~\cite{Libanov:1996vq} which did not include the complex branches of the classical background.
As we observe here, these complex branches are in fact responsible for the remarkable structure of the multiparticle amplitudes.
Let us discuss the implications in the following.

\section{Conclusions}
\label{sec:conclusions}

Calculations of high-energy processes involving the production of a large number of particles in weakly-coupled scalar quantum field theories indicate the need for novel non-perturbative behavior or even new physical phenomena.
In particular, the factorial growth of high multiplicity $1 \to n$ amplitudes at tree level, as suggested by both perturbative~\cite{Cornwall:1990hh,Goldberg:1990qk,Brown:1992ay,Voloshin:1992mz,Argyres:1992np,Smith:1992kz,Smith:1992rq} as well as semiclassical~\cite{Son:1995wz,Khoze:2017ifq,Khoze:2018kkz} methods, has sparked new ideas towards this pursuit.
A prominent example of the latter is the ``Higgsplosion" mechanism~\cite{Khoze:2017tjt}.

The approach we take in this work is more conservative.
Producing a large number of quanta in a theory with coupling $\lambda$ may be intrinsically non-perturbative for $n \gtrsim 1 / \lambda$, at weak coupling.
In this regime, the description of a tree-level process of the form $1 \to n$ by a coherent sum of Feynman diagrams may already enter the realm of large-order perturbation theory, therefore suffering from the familiar factorial growth of the coefficients of a divergent series~\cite{Dyson:1952tj}.
In this case, one would tend to believe that the rapid growth is purely a manifestation of the obvious shortcomings of a \emph{naive} approach to perturbation theory.
Hence, by including higher-order terms of the expansion, i.e.~quantum corrections to the tree-level result, we could still make physical sense of the perturbative result through a suitable resummation of the divergent series.
In particular, this is typically promising in theories where non-perturbative effects, such as instantons, are absent, which would otherwise spoil the perturbative expansion by rendering it non (Borel) resummable (see, e.g.,~\cite{Brezin:1977gk,Bogomolny:1977ty,Bogomolny:1980ur,Stone:1977au,Achuthan:1988wh,Liang:1995zq}).
Indeed, it has been argued that by including leading quantum corrections to the tree-level result, $1 \to n$ amplitudes in $\varphi^4$ theory at high multiplicities are of exponential form~\cite{Voloshin:1992nu,Khlebnikov:1992af,Libanov:1994ug,Libanov:1995gh,Bezrukov:1995qh,Libanov:1996vq},
\begin{equation}
	\mathcal{A}(n) = \mathcal{A}_{\mathrm{tree}}(n) \exp \left( \frac{F}{\lambda} \right) \, ,
\label{eq:phi4_exp}
\end{equation}
where $F$ is a function of the combination $\lambda n$ only.
In the double-scaling limit $\lambda \to 0$ and $n \to \infty$ with $\lambda n$ constant, the exponential expression for the multiparticle amplitude already indicates a form of non-perturbative behavior with $\lambda$ being in the denominator of the exponent.
Crucially, for a $\varphi^4$ theory without a spontaneously broken symmetry (i.e.~with a positive mass term), there is strong evidence that the exponent is negative, $F < 0$,\footnote{Technically, to see this, one can lift the tree-level contribution into the exponent, $F_{\mathrm{tot}} = F_{\mathrm{tree}} + F$, such that the overall exponent is negative, $F_{\mathrm{tot}} < 0$.} (see, e.g.,~\cite{Voloshin:1992nu,Libanov:1997nt} for the field theory example and~\cite{Jaeckel:2018ipq,Jaeckel:2018tdj} for its quantum mechanical analogue).
This behavior suggests that a suitable resummation of large-order perturbation theory may remedy the rapid growth of multiparticle threshold amplitudes for large $n$, at least in theories with a unique global minimum, where non-perturbative effects, such as instantons, are expected to be absent.

\bigskip

In this work, we have gone beyond $\varphi^4$ theory and computed $1 \to n$ multiparticle amplitudes at the kinematic threshold in scalar quantum field theories with higher-order self-interactions.
Closely following the methods put forward in~\cite{Libanov:1994ug,Libanov:1995gh,Bezrukov:1995qh,Libanov:1996vq}, we have calculated the quantum corrections to the matrix element $\bra{0} \varphi \ket{0}$, which generates all multipartcle threshold amplitudes~\cite{Brown:1992ay}.
By examining the leading singularity structure of the quantum fluctuations in a classical background field, we have been able to compute these quantum contributions to large orders in perturbation theory.
In turn, this enables us to determine all leading-$n$ contributions to the multiparticle amplitudes.
In the case of $\varphi^6$ theory, quite surprisingly, a complete resummation of the divergent series expansion yields
\begin{equation}
	\mathcal{A}(n) = \mathcal{A}_{\mathrm{tree}}(n) \cosh \left( \sqrt{\frac{F}{\lambda}} \right) \, .
\end{equation}
Here, schematically, $F$ is a function of the combination $\lambda^2 n^4$ only.
This is a remarkable result in its own right, because it is structurally more profound than the exponentiation found in $\varphi^4$ theory.
Nevertheless, it precisely agrees with explicit computations in the analogous quantum mechanical setting~\cite{Schenk:2019kmx}.
Crucially, it may have dire phenomenological consequences which, perhaps even more critical, are entirely independent of the exact form of the function $F$.
That is, we can immediately conclude that for $\lambda \to 0$ and $n \to \infty$ with $\lambda n^2$ constant, the multiparticle amplitude in $\varphi^6$ theory grows without bounds.
Ultimately, this may even point towards a violation of the unitarity requirement in the quantum theory.
A fully conclusive argument supporting this, however, would require a more careful calculation beyond the kinematic threshold, as we will briefly discuss momentarily.
In fact, within the framework of leading singularities, there is a straightforward generalization to a weakly-coupled scalar field theory featuring a higher-order self-interaction $\varphi^k$.
In this case, the leading-$n$ contributions to the amplitude schematically take the schematic form
\begin{equation}
	\mathcal{A}(n) = \mathcal{A}_{\mathrm{tree}} (n) \sum_{\frac{q}{2} \, \mathrm{roots}} \exp \left( \sqrt[\frac{q}{2}]{\frac{F}{\lambda}} \right) \, ,
\end{equation}
where we have defined $q = k-2$, such that $F$ is a function of the combination $\lambda^2 n^q$.
That is, remarkably, in a scalar quantum field theory with a $\varphi^k$ self-interaction (and unbroken symmetry), the multiparticle threshold amplitudes can be written as a sum over exponentials of $q/2$ complex roots of some function $F$ divided by the coupling of the theory.
Independent of the exact expression for $F$, we conclude that, apart from the $\varphi^4$ case, there will always be a piece of this combination with a positive real part, thereby dominating the rapid growth of the amplitude for large $n$ and weak coupling $\lambda \to 0$.
That means, in a generic scalar field theory, a suitable resummation of large-order perturbation theory does not remedy the rapid growth of the multiparticle amplitudes, strongly indicating a severe breakdown of perturbation theory at high energies.
At the same time, the framework itself does not provide any obvious resolution to this problem so far.
This is particularly surprising, as the mass term has been explicitly chosen to be positive in order to avoid the obvious complications associated to the presence of instantons, for example.
In this sense, $\varphi^4$ theory is an extraordinarily peculiar candidate among self-interacting scalar field theories, as it is the unique example where the intrinsic structure of the multiparticle amplitudes may allow for their asymptotic decay and thus, at least in principle, for a consistent implementation of unitarity in the high-energy regime.

However, we remark that our results do not necessarily imply a violation of unitarity in the quantum theory.
That is, the previous discussion is only valid at the kinematic threshold, where all final-state particles are at rest.
Strictly speaking, the multiparticle amplitudes are located at the edge of the physical phase space.
Therefore, the associated physical cross section vanishes, such that a rapid growth at high energies may pose no catastrophic problem.
In fact, the perturbative as well as semiclassical methods allow to move away from the kinematic threshold into a non-relativistic regime (see, e.g.,~\cite{Libanov:1994ug,Khoze:2014kka}).
Nevertheless, even in this scenario away from threshold, at least for $\varphi^4$ theory, the rapid growth of the amplitudes persists.
This property certainly calls for further investigations also in our example of theories featuring higher-order self-interactions.
It would also be interesting to examine if the inherent structure of a sum of exponential functions prevails when quantum corrections subleading in the particle number $n$ are included.

In summary, our findings indicate a possibly severe breakdown of resummed perturbation theory in the high-energy limit of scalar quantum field theories, even in scenarios where non-perturbative effects, such as instantons, are absent.
However, the inherent non-perturbative nature of the multiparticle amplitudes remains peculiar.
While the generating-field technique, only including the leading singularities of the classical background field, may not be sufficient to investigate this problem in more depth, the exact agreement with earlier, more rigorous results in the quantum mechanical analogue of scalar field theory is nevertheless striking.
We conclude that our work points increasingly towards the need for novel non-perturbative behavior or even new physical phenomena in theories of scalar quantum fields at high energies.

\section*{Acknowledgments}

I thank Joerg Jaeckel and Valya Khoze for very interesting and helpful discussions as well as for collaboration on related works.
I am also grateful to Joerg Jaeckel for valuable comments on the manuscript.
The author is funded by the Deutsche Forschungsgemeinschaft (DFG, German Research Foundation) -- 444759442.

\newpage

\bibliographystyle{inspire}
\bibliography{refs}

\end{document}